# Epitaxial Growth of Monolayer PdTe$_2$ and Patterned PtTe$_2$ by Direct Tellurization of Pd and Pt surfaces

Lina Liu,[†] Dmitry Zemlyanov[*,°] and Yong P. Chen[*,†,°,§,‖,⊥]

[†]Department of Physics and Astronomy, Purdue University, West Lafayette, Indiana 47907, USA

[°]Birck Nanotechnology Center, Purdue University, West Lafayette, Indiana 47907, USA

[§]Purdue Quantum Science and Engineering Institute and School of Electrical and Computer Engineering, Purdue University, West Lafayette, Indiana 47907, USA

[‖]WPI-AIMR International Research Center on Materials Sciences, Tohoku University, Sendai 980-8577, Japan

[⊥]Institute of Physics and Astronomy and Villum Centers for Dirac Materials and for Hybrid Quantum Materials and Devices, Aarhus University, 8000 Aarhus-C, Denmark

**ABSTRACT:** Two-dimensional (2D) palladium ditelluride (PdTe$_2$) and platinum ditelluride (PtTe$_2$) are two Dirac semimetals which demonstrate fascinating quantum properties such as superconductivity, magnetism and topological order, illustrating promising applications in future nanoelectronics and optoelectronics. However, the synthesis of their monolayers is dramatically hindered by strong interlayer coupling and orbital hybridization. In this study, an efficient synthesis method for monolayer PdTe$_2$ and PtTe$_2$ is demonstrated. Taking advantages of the surface reaction, epitaxial growth of large-area and high quality monolayers of PdTe$_2$ and patterned PtTe$_2$ is achieved by direct tellurization of Pd(111) and Pt(111). A well-ordered (2×2) PtTe$_2$ pattern with Kagome lattice formed by Te vacancy arrays is successfully grown. Moreover, multilayer PtTe$_2$ can be also obtained and potential excitation of Dirac plasmons is observed. The simple and reliable growth procedure of monolayer PdTe$_2$ and patterned PtTe$_2$ gives unprecedented opportunities for investigating new quantum phenomena and facilitating practical applications in optoelectronics.

**KEYWORDS:** *two-dimensional materials, epitaxial growth, monolayer, defect, plasmons*

## ■ INTRODUCTION

Two-dimensional (2D) noble transition metal dichalcogenides (NTMDs) are a subgroup of 2D transition metal dichalcogenides (TMDCs) which exhibit drastically different properties from most widely studied TMDCs.[1] NTMDs demonstrate strong thickness-dependent band structures, high mobility, in-plane anisotropy and long-time air stability, offering exciting opportunities in nanoelectronics, optoelectronics and catalysis.[2-9] For example, type II Dirac fermions with superconductivity in PdTe$_2$ have been shown, displaying strong possibilities for realizing the coexistence of superconductivity and topological states at 2D level for observing Majorana fermions.[10-13] For PtTe$_2$, excitations of 3D Dirac plasmons are observed, making it a promising candidate material for optoelectronic applications.[14-16] On the other hand, both PdTe$_2$ and PtTe$_2$ show thickness-dependent band structures,[17] which gives great opportunities to investigate their thickness-related electrical and optical properties. Therefore, it is highly desirable to achieve high-quality 2D PdTe$_2$ and PtTe$_2$ down to monolayers.

However, unlike common TMDCs, both PdTe$_2$ and PtTe$_2$ show strong interlayer interaction due to their specific electronic configurations (nearly full occupation of *d* orbits and high hybridization of *p* orbits),[17, 18] which makes it challenging to obtain ultrathin 2D layers through "top-down" methods such as mechanical exfoliation. Very recently, PdTe$_2$ and PtTe$_2$ few layer flakes have been grown by molecular beam epitaxy (MBE) and chemical vapor deposition (CVD),[19-21] demonstrating a promising "bottom-up" approach to synthesize monolayers. More interestingly, in another studied NTMD, platinum diselenide (PtSe$_2$), intrinsically patterned 1T/1H monolayer has been achieved during epitaxial growth by controlling concentration of defects,[22] indicating a potential way to create 2D patterns of NTMDs with designed defect arrangements.

Here, we report the first synthesis of large-area PdTe$_2$ and PtTe$_2$ monolayer films by direct tellurization of Pd(111) and Pt(111) surfaces. Using a simple one-step approach, we obtained high-quality monolayer films confirmed by scanning tunneling microscopy (STM), X-ray photoelectron spectroscopy (XPS) and low-energy electron diffraction (LEED). In the case of PtTe$_2$, we obtained a (2×2) PtTe$_2$ pattern with Kagome lattice formed by well-ordered Te vacancies on the topmost layer. Multilayer PtTe$_2$ was also obtained by this method and potential excitation of Dirac plasmons was confirmed by high-resolution electron energy loss spectroscopy (HREELS). The reliable synthesis of monolayer PdTe$_2$ and patterned PtTe$_2$ will open new application opportunities for electronics and optoelectronics.

■ **RESULTS AND DISCUSSION**

PdTe$_2$ and PtTe$_2$ monolayers were grown by deposition of Te on Pd(111) and Pt(111) surfaces followed by annealing, respectively. The schematic cartoon of the process is shown in Figure 1. A large-area PdTe$_2$ monolayer film was obtained after annealing at 470 °C (Figure 2b). LEED shows hexagonal diffraction patterns, which confirms the (1×1) PdTe$_2$ structure and indicates high crystallinity over a large area of the as-grown film (insets of Figure 2b and Figure S1a). By manually changing positions of the sample, LEED characterizations were carried out on different spots of the crystal. A series LEED results are shown in Figure S1. The diffraction patterns in all LEED images are nearly identical and displays one set of diffraction patterns, which indicates that the as-grown $PdTe_2$ film exhibits same crystalline orientation across the entire surface, proving the high quality and a large area of film with the size of several millimeters. The high-resolution STM image (Figure 2c) demonstrates the atomic structures of $PdTe_2$, which is hexagonal arrangements of Te atoms with an interatomic distance of 4.7 Å (Figure 2a and 2d). This number corresponds to ($\sqrt{3}\times\sqrt{3}$) lattice of Pd(111) ($\sqrt{3}\times 2.75=4.76$ Å), which indicates an in-plane lattice expansion of the as-grown $PdTe_2$ compared with bulk[19] due to lattice match during epitaxial growth. In the case of $PtTe_2$, a monolayer film was obtained at low annealing temperature of 200 °C except for a few individual nanoparticles on the surface (Figure 2f). Unlike the regular (1×1) structure of $PdTe_2$, LEED displays hexagonal diffraction patterns of (2×2) $PtTe_2$ (inset of Figure 2f). A unique atomic arrangement is demonstrated (Figure 2e and 2g) in which holes form hexagonal arrays with the periodicity of 8.2 Å (Figure 2h), corresponding to (2×2) $PtTe_2$ Kagome lattice. These holes are defects of Te vacancies ($V_{Te}$) due to missing of Te atoms from the topmost layer. The size of the (2×2) $PtTe_2$ structure approximately matches three times of Pt(111) lattice (2×4.1 Å ≈ 3×2.77 Å=8.31 Å). Therefore, we can suppose that the generated Te vacancies rearrange into the (2×2) pattern, which is driven thermodynamically to match the lattice of underlying Pt(111) crystal. By counting the number of atoms in the STM images, stoichiometry of the (2×2) pattern is $PtTe_{1.75}$ because one of four Te atoms on the topmost layer is missing. The (2×2) pattern demonstrates high thermal stability. Following annealing at 500 °C in ultrahigh vacuum (UHV), LEED patterns do not change and remain the (2×2) features (Figure S2a). The detailed STM image shows that the (2×2) $PtTe_2$ is still mainly intact although a few different vacancy defects are generated (Figure S2b-d). The (2×2) pattern is unique and only observed in $PtTe_2$ rather than $PdTe_2$. This phenomenon can be attributed to the higher energy of self-diffusion of atoms on Pt surface compared with Pd,[23] which leads to lower Pt-Te reactive abilities compared with Pd-Te system, resulting in losing Te easily and the formation of Te vacancies.

The thickness of the as-grown films was calculated using XPS results (Figure 3). We used the XPS Thickness Solver tool from Nanohub[24] to calculate the layer thickness.

XPS thickness model is discussed in detail elsewhere.[25, 26] This derivation follows the approach by Fadley.[27] The equation for the overlayer thickness, $t$, can be written as

$$\frac{N_l(\theta)}{N_s(\theta)} = \frac{\rho_l \times \frac{d\sigma_l}{d\Omega} \times \Lambda_l(E_l) \times \cos\theta}{\rho_s \times \frac{d\sigma_s}{d\Omega} \times \Lambda_s(E_s) \times \cos\theta} \times \frac{\left(1 - \exp\left(\frac{-t}{\Lambda_l(E_l) \times \cos\theta}\right)\right)}{\left(\exp\left(\frac{-t}{\Lambda_l(E_s) \times \cos\theta}\right)\right)} \quad (1)$$

Where $N_l(\theta)$ and $N_s(\theta)$ are the photoemission peak areas of the overlayer ($l$, Te $3d$ in our case) and substrate ($s$, Pd $3d$ or Pt $4f$ in our case) at the given photoemission angle, $\theta$, with respect to the surface normal. $\Omega$ is the acceptance solid angle of the electron analyzer. $\rho_l$ and $\rho_s$ are the numbers of atoms per unit volume (density) for the overlayer and substrate. $(d\sigma_l/d\Omega)$ and $(d\sigma_s/d\Omega)$ are differential cross-section for the overlayer and substrate photoemission peak, which can be calculated from Scofield cross sections[28] and the Reilman asymmetry parameter.[29] $\Lambda_l(E_l)$ is the electron attenuation length (EAL) of overlayer electrons attenuated in the overlayer. $\Lambda_s(E_s)$ is the EAL of substrate electrons attenuated in the substrate. $\Lambda_l(E_s)$ is the EAL of substrate electrons attenuated in the overlayer. $\Lambda_l(E_l)$, $\Lambda_s(E_s)$ and $\Lambda_l(E_s)$ are calculated using NIST SRD-82.[30] Equation (1) can be solved numerically by using Nanohub.[24] To calibrate the thickness of a monolayer NTMD, we prepared monolayer PtSe$_2$, another NTMD which exhibits same atomic structures with PdTe$_2$ and PtTe$_2$, according to the protocol in the previous study.[31] The thickness of PtSe$_2$ monolayer calculated using Equation (1) is equal to 2.26 Å. Therefore, this value is used as the reference of monolayer for our thickness calculation. The calculated thickness of epitaxially grown PdTe$_2$ and PtTe$_2$ in this study is 2.37 Å and 2.48 Å, respectively, corresponding to the thickness of monolayers. We point out that our thickness calculation based on XPS was crosschecked by transition electron microscopy (TEM) for TMDC materials such as MoS$_2$.[26]

Chemical sensitivity of XPS characterizations also helps to confirm the formation of PdTe$_2$ and PtTe$_2$. After deposition at room temperature, the centroids of the Pd $3d_{5/2}$ and Pd $3d_{3/2}$ peaks are 335.6 eV and 340.9 eV, respectively, which shift towards higher binding energies (BEs) by 0.5 eV with respect to those of Pd(111) (Figure 3a and S3a), indicating the formation of Pd-Te chemical bonds right after deposition and giving evidence of the low reaction barrier between Pd and Te. After annealing, the Pd $3d_{5/2}$ peaks shift back to 335.1 eV. This is because the contribution of Pd(111) bulk dominates the XPS signal and the contribution of PdTe$_2$ monolayer is difficult to be separated. In contrast, Pt $4f$ peaks do not show significant shift before and after annealing (Figure 3c and S3b), further indicating low reactive abilities between Pt and Te compared with Pd. The Te $3d_{5/2}$ and Te $3d_{3/2}$ peaks of PdTe$_2$ shift towards lower BEs by 0.2 eV after annealing, indicating the formation of PdTe$_2$ (Figure 3b). For both PdTe$_2$ and PtTe$_2$, the Te $3d_{5/2}$ and Te $3d_{3/2}$ peaks are 573.3 eV and 583.7 eV, respectively (Figure 3b and 3d), which are higher by ~0.5 eV than those of bulk materials.[19, 21] This could be attributed to charge transfer between the epitaxial grown monolayers and substrates due to differences in work functions.[32, 33]

The effect of annealing temperature was investigated. At low annealing temperature of 100 °C, the surfaces of PdTe$_2$ and PtTe$_2$ are not flat with islands (Figure S4a and 4a). The thickness of islands in PdTe$_2$ and PtTe$_2$ is 2.3 Å and 2.5 Å, corresponding to monoatomic steps of Pd(111) and Pt(111), respectively. The islands indicate breaking apart of monoatomic terraces and substantial mass-transport of materials during growth (Figure S4b), as well as non-thermodynamic equilibrium growth. In this case, the monolayer film behaves as a "carpet" covering the surface. Atomic structures of PdTe$_2$ can be resolved even after the low temperature annealing (inset of Figure S4a), further indicating a low reaction barrier between Pd and Te. However, in the atomic-resolution STM image of PtTe$_2$, coexistence of two structures with different atomic arrangements can be identified (labelled with "I" and "II" in Figure 4b). Structure I exhibits hexagonal atomic arrangements with the periodicity of 4.1 Å (Figure 4c and 4e), which corresponds to the (1×1) PtTe$_2$ lattice.[34] Typically, this structure is observed outside of the islands on the low areas. Structure II demonstrates the (2×2) PtTe$_2$ Kagome lattice (Figure 4d). LEED patterns display bright (1×1) and weak (2×2) diffraction spots, proving the coexistence of the two structures across a large surface area (inset of Figure 4b). Compared with the complete (2×2) pattern obtained under 200 °C, the emergence of regular (1×1) lattice of PtTe$_2$ at 100 °C is because more Te remains on the surface at low temperature. This phenomenon indicates that the amount of Te can play a significant role in the epitaxial growth of monolayer PtTe$_2$.

Therefore, we applied this method to grow PtTe$_2$ multilayers by increasing the deposited amount of Te (Figure 5a). Thickness calculated based on XPS results is 10.5 Å, corresponding to four layers approximately. As shown in Figure S5, the Pt $4f_{7/2}$ and $4f_{5/2}$ peaks of Pt are 71.6 eV and 74.9 eV, respectively, which shift to higher regions by 0.4 eV. Meanwhile, the Te $3d_{5/2}$ and $3d_{3/2}$ peaks shift to lower regions by 0.5 eV, resulting in doublet peaks at 572.8 eV and 583.2 eV. The binding energies of Pt and Te confirm the formation of PtTe$_2$.[21] According to the high-resolution STM image (Figure 5b), atomic structures were observed with the periodicity of 4.1 Å, corresponding to the lattice constant of PtTe$_2$.[34] LEED results show typical (1×1) diffraction patterns (inset of Figure 5a), indicating high crystallinity across a large surface area. Notably, potential excitation of Dirac plasmons of the multilayer PtTe$_2$ was observed at the energy of 0.4 eV by high-resolution electron energy loss spectroscopy (HREELS) (Figure 5c), which is close to the energy observed in PtTe$_2$ bulk crystal,[14] indicating high quality of the multilayer PtTe$_2$ and showing its potentials to be used in future optoelectronics. The successful growth of monolayer and multilayer PtTe$_2$ also provides a significant platform for investigating the unique layer-dependent properties of this material.

## ■ CONCLUSIONS

In summary, we demonstrate the epitaxial growth of monolayer PdTe$_2$ and patterned PtTe$_2$ by direct tellurizaion of Pd(111) and Pt(111) surfaces as confirmed by STM, XPS and LEED measurements. Large-area and high quality PdTe$_2$ and patterned PtTe$_2$ monolayer film were grown at temperature of 470 °C and 200 °C, respectively. PdTe$_2$ exhibits (1×1) lattice while PtTe$_2$ shows a (2×2) pattern which is formed by well-ordered Te vacancies. Thickness was calculated based on XPS results and monolayer features of the epitaxial-grown films were confirmed. Annealing at low temperature of 100 °C resulted in non-flat surfaces and coexistence of (1×1) and (2×2) PtTe$_2$ structures was obtained. High quality multilayer PtTe$_2$ can also be obtained by increasing the amount of deposited Te and potential excitation of plasmons at 0.4 eV of the multilayer PtTe$_2$ was measured by HREELS. The simple and robust synthesis of large area 2D materials can facilitate exploring new quantum phenomena and developing potential applications in future optoelectronics.

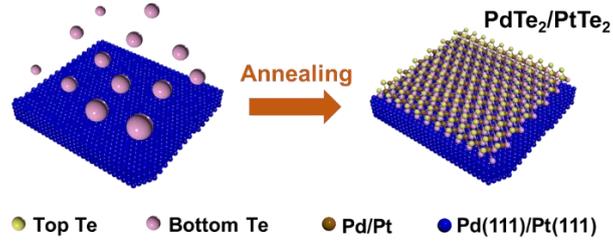

**Figure 1.** Schematic process of epitaxial growth of monolayer PdTe$_2$ and patterned PtTe$_2$ on single crystal Pd (111) or Pt (111) by tellurization.

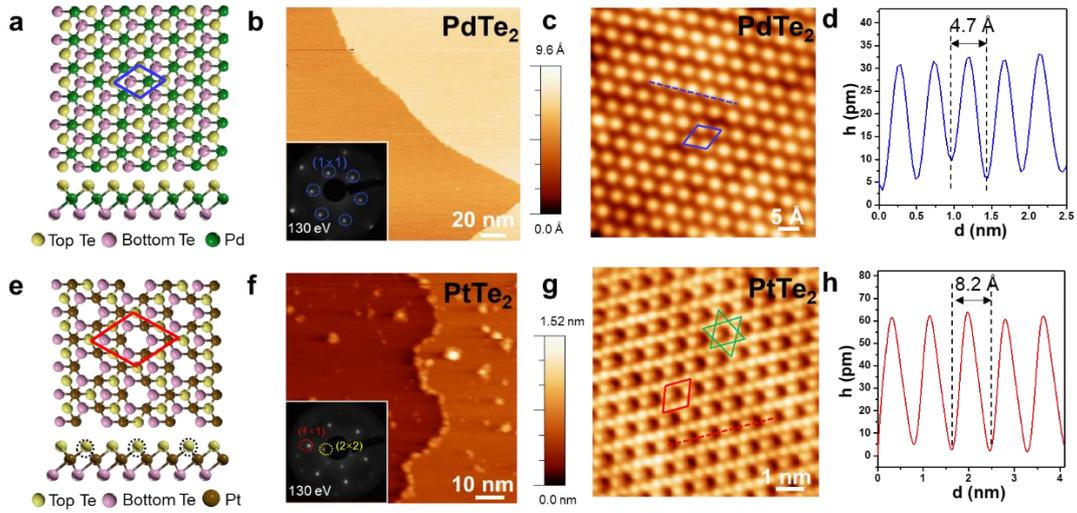

**Figure 2.** Epitaxial growth of monolayer PdTe$_2$ and patterned PtTe$_2$. (a, e) Schematics of atomic structures of as-grown PdTe$_2$ and patterned PtTe$_2$, respectively. Top view and side view. Black dashed circles in the side view in (e) indicate missing Te atoms in the atomic row. (b, f) Typical topographic STM images of as-grown PdTe$_2$ and patterned PtTe$_2$ films, respectively. Scanning parameters of (b): $V_b = 0.8$ V, $I_t = 0.7$ nA. Scanning parameters of (f): $V_b = -0.5$ V, $I_t = 1.6$ nA. Insets: LEED of PdTe$_2$ and patterned PtTe$_2$, respectively. Blue dashed circles indicate the (1×1) PdTe$_2$ lattice. Red and yellow dashed circles indicate (1×1) and (2×2) lattice of PtTe$_2$, respectively. (c, g) Atomic resolution STM images of as-grown PdTe$_2$ and patterned PtTe$_2$, respectively. Scanning parameters of (c): $V_b = 0.8$ V, $I_t = 0.4$ nA. Scanning parameters of (g): $V_b = -0.5$ V, $I_t = 0.7$ nA. (d, h) Line profiles of PdTe$_2$ and patterned PtTe$_2$ corresponding to the blue and red dashed lines in (c) and (g), respectively. Blue rhombuses in (a) and (c) represent unit cells of PdTe$_2$. Red rhombuses in (e) and (g) represent unit cells of patterned PtTe$_2$. Green triangles in (g) represent Kagome lattice in the patterned PtTe$_2$.

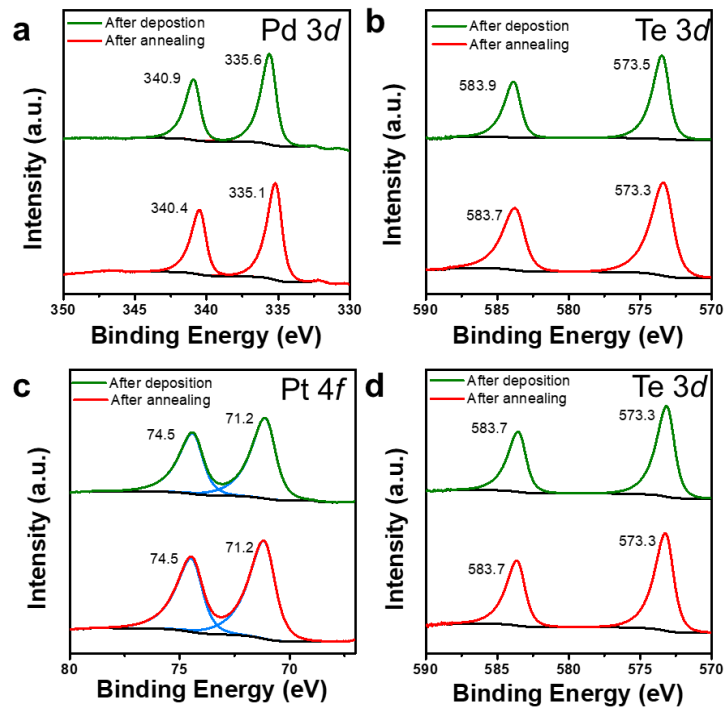

**Figure 3.** XPS spectra of as-grown PdTe$_2$ and patterned PtTe$_2$. (a, b) XPS spectra of Pd 3*d* and Te 3*d* peaks obtained after deposition of Te (green spectra) and following annealing at 470 °C (red spectra), respectively. (c, d) XPS spectra of Pt 4*f* and Te 3*d* peaks obtained after deposition of Te (green spectra) and following annealing at 200 °C (red spectra), respectively.

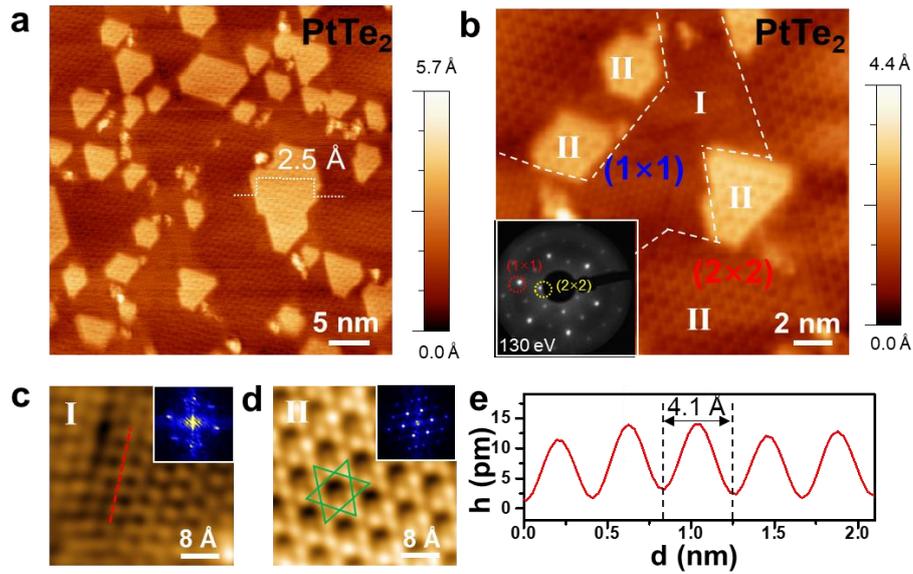

**Figure 4.** Monolayer patterned PtTe$_2$ obtained at low annealing temperature of 100°C. (a) A typical STM topographic image of PtTe$_2$. (b) A zoom-in STM image of (a). Structure I, e.g., confined by white dashed lines, exists only outside the islands. Structure II exists both inside and outside the islands. Inset: LEED patterns of the structure. Red and yellow circles indicate (1×1) and (2×2) lattices, respectively. (c, d) Atomic resolution STM images of (1×1) and (2×2) PtTe$_2$ structures, respectively. Greens triangles in (d) show the Kagome lattice. Insets of (c) and (d): Fast Fourier Transform (FFT) of each image, respectively. (e) A line profile corresponding to the red dashed line in (c). $V_b$ = -1.1 V, $I_t$ = 0.9 nA.

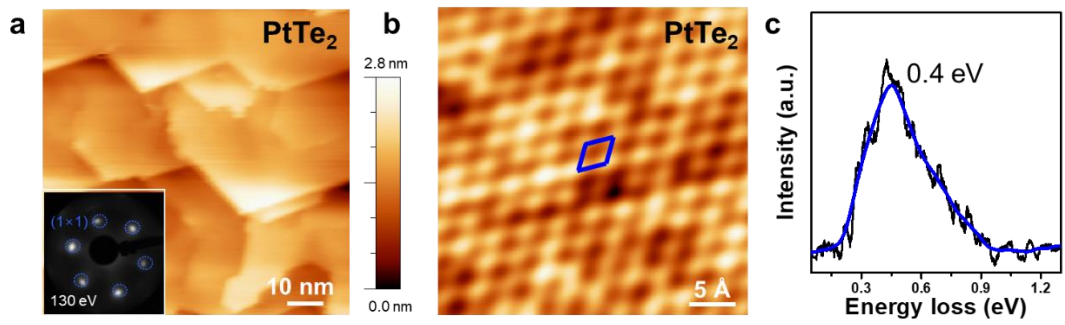

**Figure 5.** Multilayer PtTe$_2$ grown by this method. (a) A STM topographic image of multilayer PtTe$_2$. $V_b$ = -1.1 V, $I_t$ = 0.8 nA. Inset: LEED patterns of the multilayer PtTe$_2$. Blue dashed circles indicate (1×1) lattice. (b) An atomic resolution STM image of the multilayer PtTe$_2$. The blue rhombus indicates a unit cell. $V_b$ = 0.9 V, $I_t$ = 0.7 nA. (c) A HREELS spectrum of the multilayer PtTe$_2$ (after background subtraction). The blue line guides the eye. The kinetic energy of the elastic electron beam is 5 eV.

## ■ ASSOCIATED CONTENT

**Supporting Information**.

The Supporting Information is available free of charge at XXX

Experimental methods and supplementary figures (PDF)

## ■ AUTHOR INFORMATION

**Corresponding Authors**

   **Yong P Chen** - *Department of Physics and Astronomy, Purdue University, West Lafayette, Indiana 47907, USA;* Email: liu2815@purdue.edu; Email: yongchen@purdue.edu

   **Dmitry Zemlyanov** - *Birck Nanotechnology Center, Purdue University, West Lafayette, Indiana 47907, USA;* Email: dzemlian@purdue.edu

**Authors**

   **Lina Liu** - *Department of Physics and Astronomy, Purdue University, West Lafayette, Indiana 47907, USA*

**Author Contributions**

Y.C., D.Z. and L.L. conceived the experiments. L.L. and D.Z. conducted the experiments. L.L. analyzed the data and all authors discussed. L.L. wrote the paper. All authors discussed and commented on the manuscript.

**Notes**

The authors declare no conflict of interest.

## ■ ACKNOWLEDGMENTS

We acknowledge the partial financial support by the U.S. Department of Energy (Office of Basic Energy Sciences) under Award No. DE-SC0019215. We thank Prof. Ronald G. Reifenberger for fruitful discussions. We express our gratitude to the Center for Nanoscale Materials for the help with the STM tip preparation (use of the Center for Nanoscale Materials, an Office of Science user facility, was supported by the U.S. Department of Energy, Office of Science, Office of Basic Energy Sciences, under Contract No. DE-AC02-06CH11357).

# Supporting Information

# Epitaxial Growth of Monolayer PdTe$_2$ and Patterned PtTe$_2$ by Direct Tellurization of Pd and Pt surfaces

Lina Liu,[†] Dmitry Zemlyanov,[*,○] and Yong P. Chen[*,†,○,§,‖,⊥]

[†]Department of Physics and Astronomy, Purdue University, West Lafayette, Indiana 47907, USA

[○]Birck Nanotechnology Center, Purdue University, West Lafayette, Indiana 47907, USA

[§]Purdue Quantum Science and Engineering Institute and School of Electrical and Computer Engineering, Purdue University, West Lafayette, Indiana 47907, USA

[‖]WPI-AIMR International Research Center on Materials Sciences, Tohoku University, Sendai 980-8577, Japan

[⊥]Institute of Physics and Astronomy and Villum Centers for Dirac Materials and for Hybrid Quantum Materials and Devices, Aarhus University, 8000 Aarhus-C, Denmark

# 1. Experimental details

(1) Epitaxial growth of PdTe$_2$ and patternedPtTe$_2$ on Pd(111) and Pt(111)

All experiments were performed with an Omicron Surface Analysis Cluster. The system, as described elsewhere,[1-3] consists of an ultrahigh vacuum (UHV) preparation chamber and UHV $\mu$-metal analysis chamber with base pressures of $1\times10^{-9}$ mbar and $5\times10^{-11}$ mbar, respectively. Pd(111) and Pt(111) single crystals with ~9 mm of diameter, 1 mm of thickness and with orientation accuracy < 0.5° (MaTecK GmbH) were cleaned by repeated cycles of Ar$^+$ sputtering and annealing in UHV at 780 ºC. The surface cleanliness was monitored by XPS, LEED, and STM. Tellurium (99.999%, Sigma-Aldrich) was thermally evaporated using a home-built evaporator in the preparation chamber. The tellurium evaporation temperature was in the range of 290-300°C and the evaporation rate was controlled by the heater power. The amount of deposited tellurium was verified by XPS. The substrate was kept at room temperature during deposition. The sample temperature was measured by a K-type thermocouple attached to the manipulator part with which a sample holder was in a good thermal contact. For each experiment, the single crystal was freshly cleaned and Te was deposited at room temperature. Then the sample was annealed at the specified temperature for 10 mins in UHV.

(2) STM, XPS and LEED measurements

STM images (Omicron ambient temperature UHV STM) were collected at room temperature using electrochemically etched W and Pt-Ir tips at constant current (topographic) mode. In all experiments reported here, the tip was electrically grounded, meaning that at positive bias the current flowed from sample to tip. The STM images were analyzed using WSxM software.[4]

XPS was acquired using a non-monochromatic Mg K$\alpha$ X-ray source ($h\nu$ = 1253.6 eV) at 150 W. High resolution spectra were recorded at constant pass energy of 20 eV using the electron energy analyzer-Omicron EAC 125 and the analyzer controller-Omicron EAC 2000. The resolution of the instrument, which was measured as the full width at half maximum (FWHM) of Pt $4f_{7/2}$ peaks of the clean Pt(111) crystal, was approximately 1.2 eV. Photoelectrons were collected at a 45º angle with respect to the surface normal. Pt $4f$ peaks of Pt(111) were used for XPS quantification in this study. The density, kinetic energy and asymmetry parameter are 21.4 g·cm$^{-3}$, 1182.4 eV and 1.02, respectively. The calculated inelastic mean free path, transport mean free path and practical electron attenuation length (EAL) for a film thickness of 10 Å are 14.75 Å, 25.03 Å and 10.71 Å, respectively.[5]

LEED measurements were conducted in the analyst chamber straight after sample preparation with a four-grid detector (Omicron LEED spectra).

## 2. LEED measurements of as-grown PdTe$_2$

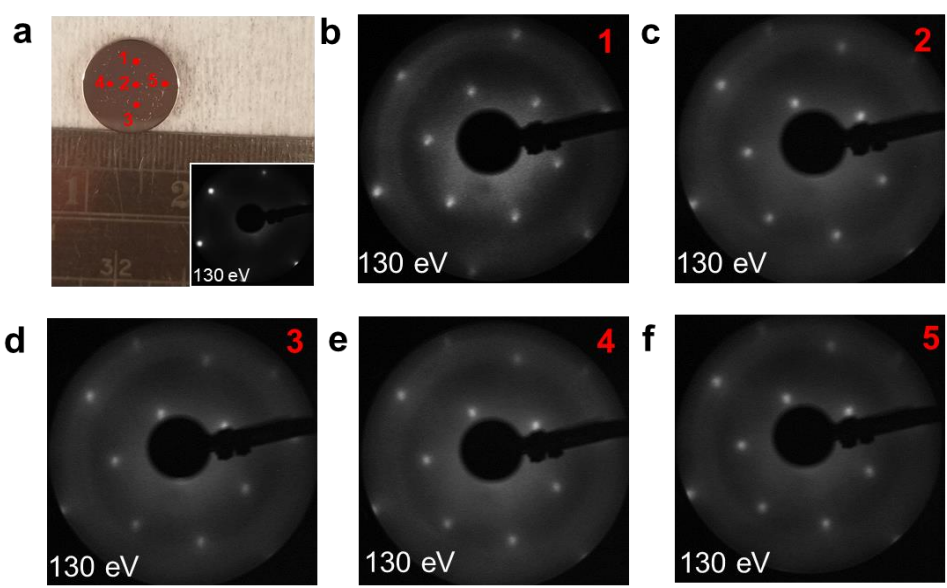

**Figure S1.** LEED measurements at different positions of the crystal. (a) A photo of the Pd(111) single crystal used in this study with a monolayer PdTe$_2$ grown on the surface. For comparison, LEED of clean Pd(111) is shown in the inset. (b-f) LEED measurements corresponding to different spots on the crystal numbered 1-5 in (a). Approximately a surface area size of ~6 mm was characterized, indicating a large PdTe$_2$ film grown on the surface.

## 3. Thermal stability of the (2×2) PtTe$_2$ pattern

After annealing in UHV at 500 ºC, LEED results of the (2×2) PtTe$_2$ pattern did not change (Figure S2a), indicating the high thermal stability of the structure. However, additional defects were observed in the detailed STM images (Figure S2b and S2c). A vacancy defect which contains one Pt atom and three Te atoms (V$_{1Pt3Te}$) was observed (Figure S2c and S2d), indicating that further missing of atoms took place locally during high temperature annealing.

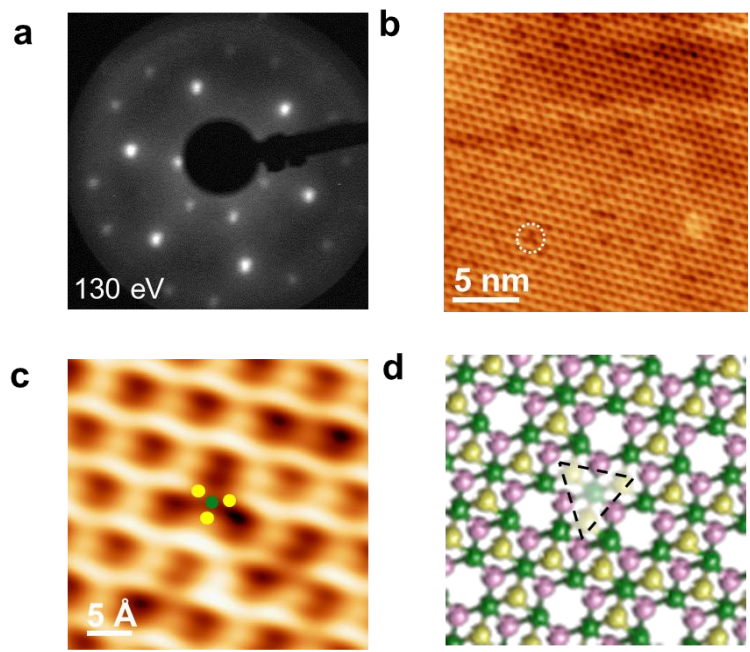

**Figure S2.** Thermal stability of the (2×2) PtTe$_2$ pattern. (a) LEED measurement of the (2×2) PtTe$_2$ pattern following annealing at 500 °C in UHV. (b) A detailed STM image of the (2×2) PtTe$_2$ pattern following annealing at 500 °C in UHV. A vacancy defect of V$_{1Pt3Te}$ is marked by the white dashed circle. (c, d) A zoom-in STM image and schematic image of the V$_{1Pt3Te}$ defect in (b), respectively. Pt and Te atoms in (c) are shown by green and yellow dots, respectively. The V$_{1Pt3Te}$ defect is marked by a black dashed triangle in (d). $V_b$ = -1.1 V, $I_t$ = 0.9 nA.

## 4. XPS spectra of clean Pd(111) and Pt(111)

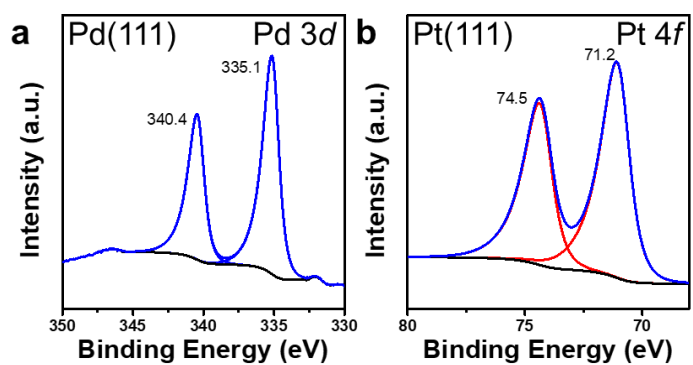

**Figure S3.** (a, b) XPS Pd 3$d$ and Pt 4$f$ spectra obtained on clean Pd(111) and Pt(111), respectively.

## 5. STM images of PdTe$_2$ obtained at low temperature and clean Pd(111)

The STM image of clean Pd(111) exhibits flat and large terraces with the width of tens of nanometers (Figure S4b), which is dramatically different from the small and irregular islands after the growth of PdTe$_2$ in Figure S4a, indicating breaking apart of terraces and mass-transport during growth.

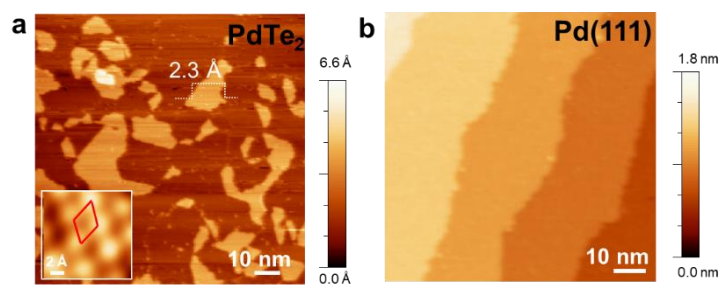

**Figure S4.** (a) A typical STM topographic image of PdTe$_2$ obtained at 100 °C. Inset: Atomic resolution STM image of PdTe$_2$. The red rhombus represents a unit cell. $V_b$ = -1.3 V, $I_t$ = 0.6 nA. A typical STM topographic image of clean Pd(111). $V_b$ = -0.9 V, $I_t$ = 0.7 nA

## 6. XPS characterizations of multilayer PtTe$_2$

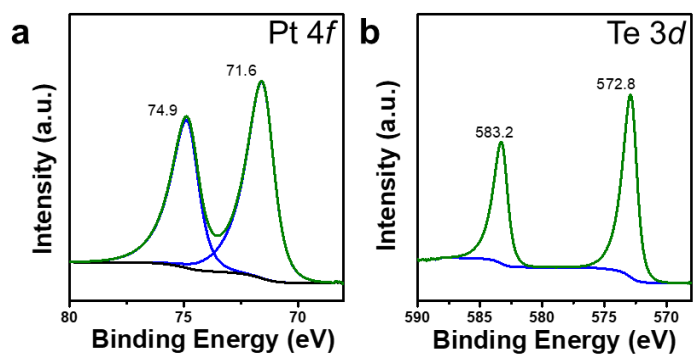

**Figure S5**. XPS spectra of multilayer PtTe$_2$. (a, b) Pt 4$f$ and Te 3$d$ peaks of multilayer PtTe$_2$, respectively.